# A simplification design of the fiber cavity ring-down technique

SUSANA SILVA,[1] ORLANDO FRAZÃO[1,*]

[1]*INESC TEC – Instituto de Engenharia de Sistemas e Computadores - Tecnologia e Ciência, Rua do Campo Alegre 687, 4169-007 Porto, Portugal,*
*\*Corresponding author: ofrazao@inesctec.pt*

**This paper presents a simplification design for the fiber cavity ring-down technique. The new configuration is no longer in ring cavity and is converted to a linear cavity. The proposed linear cavity is created with a single fibre coupler of 99:1 and two thin-film mirrors located at the end of the fibre arms. This configuration can be interrogated in reflection or in transmission (whether if two fibre couplers are used in conventional configuration), however the reflection interrogation presents better resolution when compared signal-to-noise ratio in the decay time measurement. The mirrors can be also replaced by fibre Bragg gratings and fibre loop mirrors. One of the main advantages of this simplification is its use for remote sensing. However, it is necessary to include amplification in the cavity to eliminate fiber length losses. This configuration allows using any intensity-based optical fibre sensor in the literature with twice improvement in the sensitivity and can be easily adapted to a commercial system. A proof-of-concept for strain sensing is demonstrated. This simplification opens new paths in physical, chemical or biological parameters using this simple fibre configuration.**

**Introduction.** The cavity ring-down (CRD) technique comprises two main configurations that are the basis of the known state-of-the-art: The first one uses two highly reflective mirrors thus forming a linear cavity [1], while the second one uses two fibre couplers with high splitting ratios that form a cavity ring [2]. In the first configuration, a light impulse is sent into the cavity and travels inside it by multiple internal reflections. The part of light that comes out in each turn is monitored at the output. The result is a decaying exponential behaviour of the signal's intensity with time. In the second configuration, the impulse is sent into the cavity and rings around at certain amount of time. In each loop, the intensity of the impulse decays with time and, in a similar way, an exponential behaviour is observed at the system's output.

The cavity ring is the most common configuration used in fibre-based CRD. Several schemes have already been reported for sensing [3,4]. An improvement of this technique was the use of the optical time domain reflectometer (OTDR) for sending the modulated light into the cavity ring. In 2014, this configuration was reported as a proof-of-concept for displacement sensing [5].

The OTDR is a commercial device widely used for measuring losses along several kilometres of optical fibre [6]. Early it was shown to be a promising device in measuring point-by-point losses, by using intensity sensors along the fibre. Several works have been reported in this area of research, where an OTDR is used to monitor sensors such as fibre Bragg gratings (FBGs) [7], long period gratings (LPGs) [8], multimode interference (MMI) [9], fibre loop mirrors [10] and others. When integrated in a CRD configuration, the OTDR has shown to be an excellent tool to be used as both input and output device. In 2015, the CRD technique was presented for strain sensing, where a chirped-FBG was used as sensing element and the signal's output was monitored by an OTDR [11]. Later in that year, a similar configuration was reported for measuring curvature, but in this case, a LPG was used instead as sensing device [12].

In 2016, a new configuration for remote sensing was demonstrated [13]. The cavity ring was modified with an inclusion of an optical circulator and erbium doped fibre amplifier (EDFA). In this case, a thin-film mirror was insert at the end of 20 km of fibre. The main limitation of this configuration is the intrinsic high losses of the optical circulator that obliges applying an EDFA in the cavity ring in order to compensate them.

In this work, a simple fibre linear CRD design is proposed. The new configuration consists in a linear cavity using a single fibre coupler and two thin-films mirrors located at the end of the fibre arms. It is also intended to demonstrate the viability of this new CRD technique using different types of mirrors such as the fibre loop mirror and fibre Bragg grating. Remote sensing is also demonstrated, using amplification in the linear cavity, in a maximum of 10 km. As a proof-of-concept, the linear cavity demonstrator relies on using a FBG as mirror and simultaneously as strain sensor.

**Experimental setup and results.** The schematic of the proposed fibre linear CRD configuration is presented in Figure 1. The conventional CRD configuration is also represented for better understanding the innovation of the proposed linear design. Such configuration consists on a single fibre coupler, with high splitting ratio, and two thin-film mirrors located at the end of ports 2 and 4. When compared with the conventional CRD, this linear configuration eliminates the output coupler, which is replaced by two highly reflective mirrors, as schematically shown in Figure 1.

The uniqueness of this CRD configuration is that a single fibre coupler with highly reflective mirrors at the end of the fibre arms form the linear cavity. In addition, the output signal is read in reflection, in port 1. Therefore, the oscilloscope (and photodetector) at the output (port 3) is no longer needed to interrogate the sensing head, as an OTDR serves that purpose. This is one of the major advantages of the proposed linear CRD design, combining the fact that the output signal acquired by the OTDR provides measurements in dB, which allows attaining the decay time ($\mu$s) with a linear response (rather than an exponential behaviour obtained by the oscilloscope). Further, an increase of the sensitivity is achieved because the light passes twice by the sensor [13]. Another unique feature is that que mirrors can be based on highly reflective thin-films, fibre loop mirrors or even FBGs, as it will be shown in experimental results.

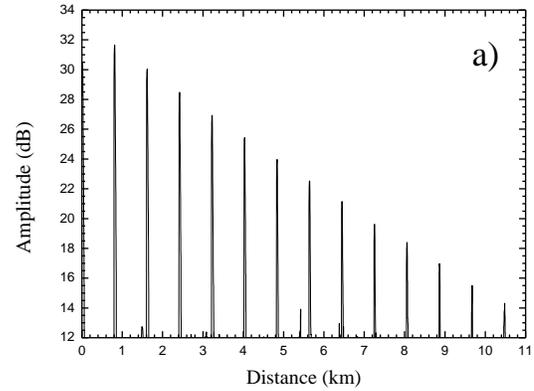

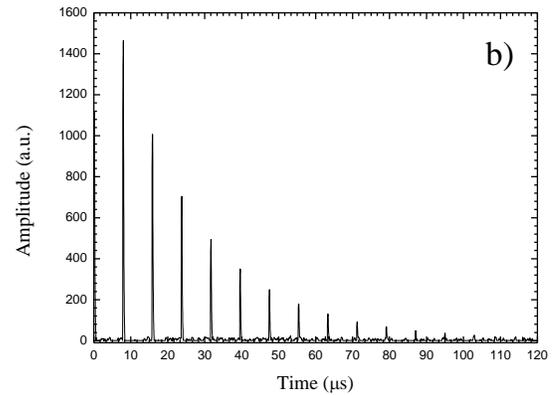

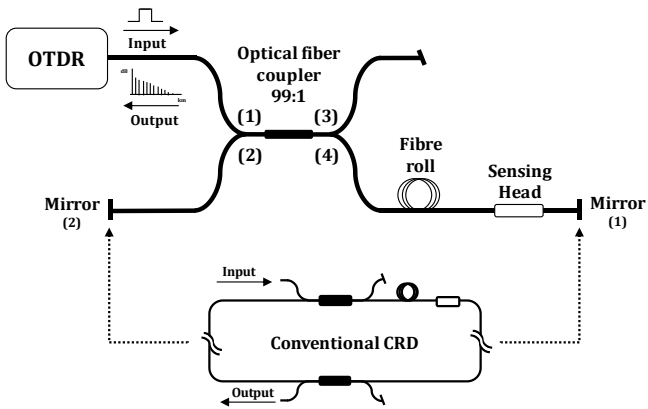

Fig. 1. Schematic of the simplified linear CRD design based on the conventional CRD configuration.

Fig. 2. a) Trace observed on the OTDR for a fibre roll with ~800 m used in port 4 (Figure 1); b) trace after signal processing.

The basis of such configuration relies on the following operation principle: the OTDR (centred at 1550 nm) is placed in port 1 and send pulses into the linear cavity, which is formed by an optical fibre coupler with a splitting ratio of 99:1. Each pulse enters in the cavity by means of 1% arm of the fibre coupler (port 4) and is back-reflected in mirror 1; then, passes through 99% arm of the fibre coupler (port 2) where is back-reflected again in mirror 2. In order to use this linear CRD configuration for sensing, a sensor head is placed in the arm of port 4, before mirror 1. In this manner, the amplitude of the pulses will slowly decay as it travels inside the linear cavity, similar to the behaviour of the conventional CRD using a cavity ring.

In order to validate this new concept, highly reflective mirrors based on a gold thin-film (50 nm) were placed at the end of ports 2 and 4. For the proof-of-concept, the OTDR was set with an operation wavelength at 1550 nm and it was used to send pulses with 200 ns width into the linear cavity. A fibre roll with ~800 m was also used in port 4 (see Figure 1).

Figure 2a) presents the trace directly observed on the OTDR, while Figure 2b) presents the trace after signal processing. The decay time of the proposed linear CRD is 21.47 $\mu$s (Figure 2b).

The linear CRD was also demonstrated for remote sensing. In order to use the proposed configuration for such purpose, a roll of fibre with ~10 km in length was placed in port 4, before the sensor head and mirror 1. This length of fibre was chosen due to the limitation of the OTDR to its reading, i.e., the maximum distance range setting is 100 km. The trace for larger lengths will not be observed due to the reduced number of acquired peaks, turning difficult the measurement of the decay time. A bidirectional amplifier was also required in for remote sensing. In this case, an Erbium Doped Fibre amplifier (EDFA) was inserted in the linear cavity to provide an observable signal at the output, eliminating the length-induced losses.

Figure 3 shows the trace acquired by the OTDR, after signal processing, for a fibre roll with ~10 km in length. Traces are depicted by the linear response (a) and the corresponding exponential behaviour (b).

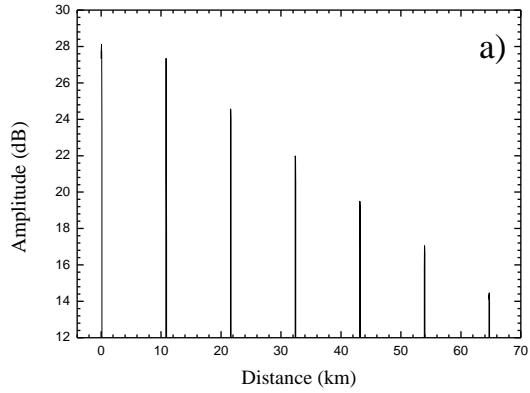

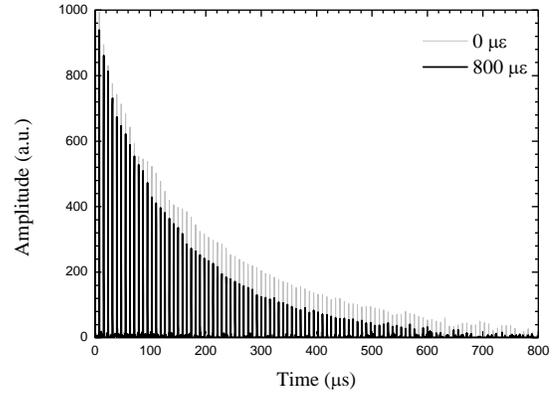

Fig. 4. Output trace after signal processing for 0 με applied to the sensor and 800 με of applied strain.

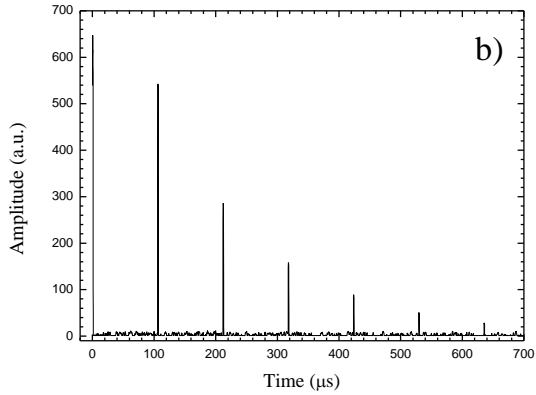

Fig. 3. a) Trace observed on the OTDR for a fibre roll with ~10 km used in port 4 (Figure 1); b) trace after signal processing.

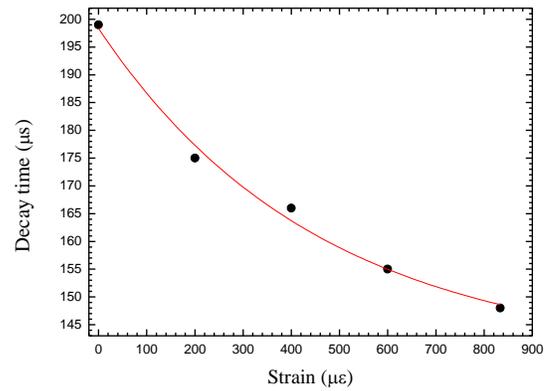

Fig. 5. Relationship between the decay time and applied strain to the chirped-FBG.

The spacing between consecutive peaks increases with increasing fibre length. In this case, the decay time was also determined and a value of 165.21 μs was attained. A linear behaviour of the decay time with increasing fibre length is expected and a slope of 14.09 μs/km was estimated.

For demonstrating sensing features of the developed new linear CRD concept, a FBG was used as mirror and simultaneously as a sensor in port 4 (see Figure 1). In this case, a chirped-based FBG was used due to its large bandwidth, acting as a band-rejection filter in reflection. The chirped-FBG placed inside the linear cavity is centred at 1554 nm and it has ~4 nm width. The output signal is back-reflected by mirror 2 (gold thin film) and interrogated by the OTDR.

To perform strain measurements, the chirped-FBG was fixed at two points that were 300 mm apart, and submitted to specific strain values by means of a translation stage (via sequential 20 μm displacements). Figure 4 shows the trace at the output, after signal processing, for the cases of 0 με applied to the sensor and 800 με of applied strain. By performing the exponential fit, it was possible to determine a decay time of 199.12 μs and 148.8 μs for 0 με and 800 με of strain applied to the sensor, respectively.

Figure 5 presents the response of the FBG when submitted to strain within the range [0-800] με. A non-linear behaviour is observed due to the spectral response of the FBG being located in a non-linear region of the multimode laser source optical spectrum. This configuration presents high sensitivity to strain when compared with the literature [11]. For a small range of applied strain, [0-200] με (highest sensitivity range, see Figure 5), a sensitivity of 120 ns/με was obtained. The proposed linear CRD configuration presents a sensitivity 90-fold the one obtained in [11].

**Conclusions.** This work proposes a new concept of the fibre CRD technique based on a linear configuration. The new approach consists in a linear cavity using a fibre couple and two thin-film mirrors located at the end of the fibre arms. The mirrors can be replaced by fibre Bragg gratings and fibre loop mirrors. The linear CRD was also demonstrated for remote sensing in a maximum of 10 km of fibre length. As a proof-of-concept for sensing, the linear cavity demonstrator relied on using a chirped-FBG as mirror and simultaneously as strain sensor. The uniqueness of this new linear CRD configuration is the use of a single fibre coupler with two highly reflective mirrors; the use of the OTDR as both input and output device; and the use of different types of mirrors that can be used as sensor elements, thus featuring remote sensing in a compact and simple configuration. When compared to the conventional fibre CRD, the simplicity of the configuration is a significant feature, as the elimination of one of the couplers allows acquiring the reflected output signal and the elimination of splices between couplers.

**Acknowledgments.** Project CORAL – Sustainable Ocean Exploitation: Tools and Sensors (NORTE-01-0145-FEDER-000036), financed by the North Portugal Regional Operational Programme (NORTE 2020), under the PORTUGAL 2020 Partnership Agreement, and through the European Regional Development Fund (ERDF); Susana Silva received a Post-Doc fellowship from FCT – Portuguese national funding agency for science, research and technology (SFRH/BPD/92418/2013).